\documentclass[12pt,onecolumn,final]{IEEEtran}
\newlength \figwidth
\usepackage{graphicx}
\usepackage{amsmath}
\usepackage[noadjust]{cite}
\usepackage{amsfonts}
\usepackage{amssymb}
\usepackage{subfigure}
\usepackage{pgfplots}
\usepackage{mathrsfs}
\usepackage{algorithm,caption}
\usepackage{algpseudocode}
\usepackage{xcolor}

\usepackage{array}
\usepackage{booktabs}

\DeclareMathOperator*{\argmax}{argmax}

\begin{document}

\newtheorem{definition}{Definition}
\newenvironment{definitionbox} {\begin{definition}}{\hfill \interlinepenalty500 $\Box$\end{definition}}
\newtheorem{condition}{Condition}
\newenvironment{conditionbox} {\begin{condition}}{\hfill \interlinepenalty500 $\Box$\end{condition}}
\newtheorem{example}{Example}
\newenvironment{examplebox} {\begin{example}}{\hfill \interlinepenalty500 $\Box$\end{example}}
\newtheorem{lemma}{Lemma}
\newenvironment{lemmabox} {\begin{lemma}}{\hfill \interlinepenalty500 $\Box$\end{lemma}}

\title{Enhanced Multiuser Superposition Transmission through Structured Modulation}

\author{\IEEEauthorblockN{{Dong~Fang, Yu-Chih~Huang, Giovanni~Geraci, Zhiguo~Ding, and Holger~Claussen}}
\thanks{D.~Fang is with UnitedHealth Group Inc., Ireland. Y.-C.~Huang is with the National Taipei University, Taiwan. G.~Geraci is with Universitat Pompeu Fabra, Spain. Z.~Ding is with University of Manchester, UK. H.~Claussen is with Nokia Bell Labs, Ireland.}
\thanks{The material contained in this paper has been in part presented at IEEE Globecom 2016 \cite{FanHuaDin2016} and at IEEE VTC Spring 2017 \cite{GerFanCla2017}.}
}

\maketitle
\thispagestyle{empty}
\vspace*{-0.5cm}
\begin{abstract}
The 5G air interface, namely, dynamic multiple access (MA) based on multiuser superposition transmission (MUST) and orthogonal multiple access (OMA), may require complicated scheduling and heavy signaling overhead. To address these challenges, we propose a a unified MA scheme for future cellular networks, which we refer to as \textit{structured multiuser superposition transmission} (S-MUST). In S-MUST, we apply complex power allocation coefficients (CPACs) over multiuser legacy constellations to generate a composite constellation. In particular, the in-phase (I) and quadrature (Q) components of the legacy constellation of each user are separately multiplied by those of the CPACs. As such, the CPACs offer an extra degree of freedom for multiplexing users and  guarantee fairness in symmetric broadcast channels. This new paradigm of superposition coding allows us to design IQ separation at the user side, which significantly reduces the decoding complexity without degrading performance. Hence, it supports low-complexity frequency-selective scheduling that does not entail dynamically switching between MUST and OMA. We further propose to quantize the CPACs into complex numbers where I and Q components of each quantized coefficient are primes, facilitating parallel interference cancellation at each user via modulo operations, last but not least, we generalize the design of S-MUST to exploit the capabilities of multiantenna base stations. The proposed S-MUST exhibits an improved user fairness with respect to conventional MUST (134$\%$ spectral efficiency enhancement) and a lower system complexity compared with dynamically alternating MUST and OMA.
\end{abstract}
\IEEEpeerreviewmaketitle
\section{Introduction}

The key performance indicators (KPI) of the fifth generation (5G) cellular networks include massive device connectivity, high data rates, ultra-high link reliability, and low energy consumption \cite{Nokia:15,Ericsson:16}. To meet the asserted KPIs, new air interfaces which may include a new paradigm of multiple access {(MA)} are called for. Compared with orthogonal multiple access (OMA) schemes, non-orthogonal multiple access (NOMA) offers superior spectral efficiency with less processing overhead, and hence has received considerable attention \cite{SaiKisBejVTC2013,BenSaiKis2013,KimLimKim2013MILCOM,SunHanI2015,LiuWan2016}. In NOMA, lower and higher powers are allocated to nearer and farther users (UEs), respectively, enabling to schedule multiple users in the same physical resource such as time/frequency/code/space \cite{Ding_comms_mag,Ding_MIMO_NOMA}. At the farther user side, the desired signal is directly decoded by treating the nearer user's signal as noise. At the nearer user side, the farther user's signal is decoded, reconstructed, and subtracted from the received signal first; then, the desired signal is decoded.

Since the 87-th meeting of the 3rd Generation Partnership Project (3GPP), NOMA has been selected as a study item in long-term evolution (LTE) release 13 -- termed multiuser superposition transmission (MUST) \cite{3GPP36859} -- and further selected as a work item in LTE release 14. MUST is classified into 3 categories: (i) in Cat.~1, each user maps its data onto a component constellation which is adaptively allocated power levels based on the near-far geometry. The composite constellation employs non-Gray mapping; (ii) in Cat.~2, the composite constellation employs Gray mapping where the adaptive power allocation and legacy mapping of each user's data are jointly designed; (iii) in Cat.~3, the composite constellation retains the legacy uniform quadrature amplitude modulation (QAM) with Gray mapping and without adaptive power allocation for users. The nearer and farther users' data is protected by unequal error protection in terms of minimum Euclidean distances. {The pros and cons of MUST Cat. 1-3 are summarized in Table \ref{tab:pros_cons}.
\begin{table}[]
\centering
\caption{\label{tab:pros_cons}Summary of pros and cons of 3 categories of MUST}
\begin{tabular}{p{3cm}<{\centering}p{4cm}<{\centering}p{4cm}<{\centering}}
 \toprule
 & Pros & Cons \\ 
 \midrule
MUST Cat. 1 & Amplitude-weighted superposition; high spectral efficiency & non-Gray labeled; cannot use legacy constellation at BS side \\ \hline
MUST Cat. 2 & Amplitude-weighted superposition; high spectral efficiency,Gray labeled & cannot use legacy constellation at BS side \\ \hline
MUST Cat. 3 & Bit-level superposition; high spectral efficiency,Gray labeled & no adaptive power allocation \\
\bottomrule
\end{tabular}
\end{table}}
As the nature of MUST is superposition coding, it only outperforms OMA in asymmetric Gaussian broadcast channels \cite{Cov1972}. Indeed, in symmetric broadcast channels, MUST Cat. 1-3 struggles to provide good user fairness \cite{TimKri2015,LiuElkDin2016}. For example, in the case of two users and in order to avoid overlapping points on the composite constellation, MUST may not be able to assign equally strong power to both users. As a result, in symmetric broadcast channels where both users should be served with equal transmission rates, the rate of one user may be higher than that of the other. In a practical design, it is the duty of the scheduler to pair users with a near-far geometry in order to retain the spectral efficiency gain of MUST \cite{Ding_SigAlgn}. However, such scheduler results in an excessively high complexity if the cellular network has a high user density and traffic demand, as the BS needs to exhaustively search through all users and pair those satisfying the condition of asymmetrical broadcast channel. While a frequency selective scheduler could allow the co-existence of OMA and MUST (dynamic MA), it would need to compare the proportional fairness (PF) metric of MUST and OMA and align the best transmission (Tx) mode across all sub-bands. Such high computational burden limits the use of MUST in massive connectivity scenarios. 
Motivated by the aforementioned challenges and practical issues of MUST, it is necessary to design an efficient downlink superposition transmission scheme that provides: (i) a unified air interface not requiring to dynamically switch between two MA schemes; (ii) a low complexity scheduler; and (iii) good user fairness in symmetric broadcast channels.

This paper aims at overcoming the aforementioned limitations of MUST with a new \emph{structured multiuser superposition transmission} (S-MUST) scheme. S-MUST employs complex power allocation coefficients (CPAC) over the in-phase (I) and quadrature (Q) components of the multiple users' legacy constellations to generate a composite constellation. As such, the CPACs offer an extra degree of freedom to guarantee user fairness in symmetric channels. The proposed S-MUST results in a unified air interface capable of replacing dynamic MA -- i.e., the alternation of MUST and OMA -- thus reducing the complexity of the frequency selective scheduler. We also quantize the CPACs into complex numbers where I and Q components of each CPAC are primes, enabling {modulo operations based parallel interference cancellation (M-PIC)} with respect to these primes, at UE side. Such M-PIC operation can be performed independently at each UE and irrespective of other users' network assistance information, such as modulation and coding scheme (MCS), power level, channel quality indicator (CQI), and precoding matrix index (PMI) -- hence significantly reducing the signaling overhead. The main contributions of this paper are three-fold and can be summarized as follows:
\begin{itemize}
\item A new non-orthogonal multiuser superposition transmission scheme, S-MUST, is proposed.
\item We provide composite constellation and mapping design for proposed S-MUST. A detection algorithm as well as the assignment of the complex power coefficients accounting for the user fairness optimization are devised.
\item We design low-complexity frequency-selective scheduling and pairing algorithms for S-MUST.
\item We extend the design of S-MUST to exploit the capabilities of multiantenna base stations (BSs), through a framework based on user selection, clustering, and zero forcing beamforming.
\end{itemize}

{The structure of this paper is as follows: 1) the challenges of existing schemes and the motivations of our design are introduced in Section II; 2) the detailed design is shown in Section IV, including transmission and reception; power allocation; user fairness protection and scheduling; 3) the joint design of MIMO and S-MUST is discussed in Section IV; 4) the performance evaluation is provided in Section V; and 5) conclusive remarks are given in Section VI.}   
\section{Challenges and Motivations}
{In this section, we present some preliminaries of conventional MUST and of dynamic MA. We then discuss the high scheduling complexity of existing schemes and the user fairness issue.}


\subsection{High scheduling complexity} 
In order to implement a dynamic MA scheme, a frequency-selective scheduler is required to select the best Tx mode -- opportunistically alternating between MUST and OMA -- for each UE across each sub-band \cite{R1-156107,R1-155931}, based on a PF metric:
\begin{equation}
{\text{P}}{{\text{F}}_\ell } = \sum\limits_{\ell  \in \mathcal{U}} {\frac{{{R_\ell }\left[ {t,\mathcal{U}} \right]}}{{{{\bar R}_\ell }\left[ t \right]}}},
\end{equation}
where ${{R_\ell }\left[ {t,\mathcal{U}} \right]}$ denotes the instantaneous rate of UE $\ell$ at time $t$ (the time index of a subframe); {${{\bar R}_\ell }\left[ t \right]$} denotes the average rate of UE $\ell$; and {$\mathcal{U}$ is the set of UE indices}. In the multiuser case, e.g., with two UEs, the PF metric, denoted by $\text{PF}_{j,k}$ ({where $j$ and $k$ are indices of paired UEs}), can be calculated from the ratio of the paired UEs' instantaneous sum-rate over their average sum-rate. PMI and CQI feedback is required to evaluate the channel condition of each sub-band. The power coefficients to the farther UE, denoted by $\alpha$, are determined in the MUST scheduling loop. {This kind of scheduling is channel dependent, commonly used in cellular systems, and referred to frequency-selective scheduling.  Instead of exploiting the frequency diversity of the channel, frequency-selective scheduling leverages the channel’s time and frequency selectivity to allocate valuable radio resources in an optimal manner.} The main frequency selectivity scheduling operations of MUST are summarized in {Algorithm 1} \cite{R1-156107,R1-155931}. In order to dynamically switch between MUST and OMA and retain the gain provided by MUST, the above scheduling algorithm requires to traverse all sub-bands several times to exhaustively search for the best Tx mode. The computational complexity thus increases exponentially with the number of sub-bands and UEs.

%
%

\subsection{User fairness loss} In addition to the complexity of a frequency-selective scheduler, another drawback of MUST is the inability to guarantee user fairness in symmetric broadcast channels. In a superposition transmission scheme, user fairness is defined as the maximum rate of the weakest UE across all sets of paired UEs \cite{TimKri2015}, given by:
\begin{equation}
\begin{aligned}
  &\mathop {\max }\limits_{{\alpha}} \mathop {\min }\limits_{i \in \mathcal{U}_s} {R_\ell }\left( \alpha  \right) \\
  &\text{s.t.}\sum\limits_{\ell  = 1}^L {{\alpha _\ell }}  \le P,0 \le {\alpha _\ell },
\end{aligned}
\end{equation}
where $R_\ell(\alpha)$ denotes the $\ell$-th UE's rate, $\alpha$ denotes the power coefficient satisfying the power constraint, and $\mathcal{U}_s$ denotes the set of paired UEs.

\begin{figure}[!htp]
\centering
\includegraphics[width=\figwidth]{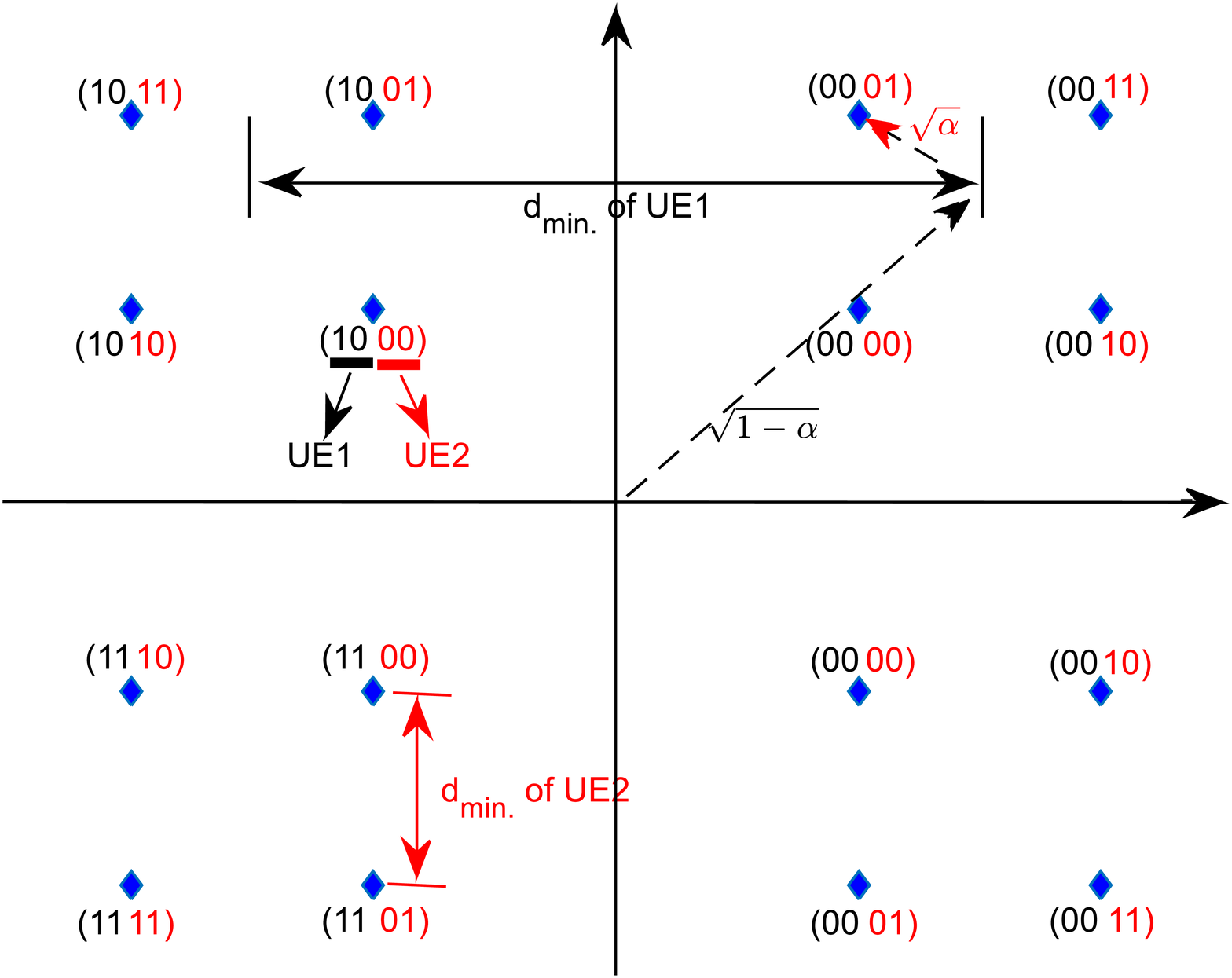}
\caption{Composite constellation of MUST Cat. 2 \cite{3GPP36859}.}
\label{fig_Const_MUST}
\end{figure}

{In the following, we will discuss why standard MUST cannot guarantee user fairness by taking MUST Cat. 2 as an example. Fig. \ref{fig_Const_MUST} illustrates the composite constellation of MUST Cat. 2., where there are far and near UEs, denoted by UE 1 and UE 2, respectively. Suppose both UEs adopt 4-ary constellation, namely, 2 bits/symbol rates. In MUST Cat. 2, the Gray mapped 16QAM is virtually treated as the superposition constellation so that each symbol on it can be treated as a superimposed symbol of both users. The first 2 bits are assigned to near UEs, marked by black. The last 2 bits are assigned to far UE, marked by red. As such, one can observe that the minimum Euclidean distance of far UE is larger than that of near UE. This indicates the far UE has higher error protection.} In symmetric broadcast channels where both UEs should be served with the equal rates, in order to avoid an overlap on the composite constellation, two UEs will be assigned with different powers, i.e., different error protection in terms of minimum Euclidean distance. As a result, user fairness cannot be guaranteed, especially in low-to-moderate signal-to-noise ratio (SNR) regimes. As such, the scheduling algorithm of MUST attempts to pair users with asymmetric channels (see the condition $\text{CQI}_j>\text{CQI}_k$ in Algorithm 1, ``UE pair selection for MUST").

\begin{algorithm}
\caption*{\textbf{Algorithm 1: Dynamic MA Frequency-Selective Scheduling}\,\cite{R1-156107,R1-155931}}
\begin{algorithmic}[1]
\State Given PMI and CQI feedback and the range of $\alpha$: $(0.025, 0.3]$;
\State Initialize the set of paired UEs ${\mathcal{U}_s}=\emptyset$;
\State \textit{Single UE selection for OMA:}
\For {each sub-band}
\For {each UE $i$ in the active UE set $\mathcal{U}$}
  \State calculate PF$_i$;
\EndFor
\item $\hat i = \mathop {\arg \max }\limits_{i \in \mathcal{U}} \{ {\text{P}}{{\text{F}}_i}\}$;
\item ${\mathcal{U}_s} \leftarrow {\mathcal{U}_s} \cup \left( {\hat j,\hat k} \right)$
\EndFor
\State \textit{UE pair selection for MUST:}
\For {each sub-band}
 \For {each near-far UE pair $\left( \text{UE}_j,\text{UE}_k \right)$ in $\mathcal{U}$}
 \If {PMI$_j$=PMI$_k$ \textbf{and} $\text{CQI}_j>\text{CQI}_k$};
 \For {all $\alpha$}
 \State $\hat \alpha  = \mathop {\arg \max }\limits_{\alpha} \left\{\text{PF}_{j,k}(\alpha)\right\}$;
 \State calculate PF$_{j,k}({\hat\alpha})$;
 \EndFor
 \Else
 \State continue;
 \EndIf
   \EndFor
\State $\left( {\hat j,\hat k} \right) = \mathop {\arg \max }\limits_{i \in \mathcal{U}} \{ {\text{P}}{{\text{F}}_{j,k}}\left( {\hat \alpha } \right)\}$;
\State ${\mathcal{U}_s} \leftarrow {\mathcal{U}_s} \cup \left( {\hat j,\hat k} \right)$.
 \EndFor
\end{algorithmic}
\end{algorithm}

\begin{algorithm}
\caption*{\textbf{Algorithm 1: Dynamic MA Frequency-Selective Scheduling}\,\cite{R1-156107,R1-155931} (continued)}
\begin{algorithmic}[1]
\setcounter{ALG@line}{25}
\State \textit{Tx mode selection:}
\For {each sub-band}
\If {${\text{P}}{{\text{F}}_{\hat j,\hat k}}\left( {\hat \alpha } \right)>{{\text{PF}}_{\hat i}}$}
\State Tx mode$=$MUST;
\Else
\State Tx mode$=$OMA.
\EndIf
\EndFor
\State \textit{UE alignment and sub-band release:}
\For {each UE}
\If {no. of MUST Tx. mode$>$no. of OMA Tx. mode}
\State best Tx mode$=$MUST;
\Else
\State best Tx mode$=$OMA;
\EndIf
\State Release sub-bands where selected UE is scheduled with other Tx mode than the best Tx mode. UE selected is such sub-bands must be scheduled with the best Tx mode in the next scheduling round.
\EndFor
\end{algorithmic}
\end{algorithm}

\section{Detailed Design of S-MUST}

{In this section, we provide a detailed design for the proposed S-MUST scheme, including transmission design in subsection A, reception design in subsection B, power allocation in subsection C, user fairness in subsection D and scheduler design in subsection E. The main advantage of the proposed S-MUST over a hybrid MA scheme (OMA and MUST) is that S-MUST does not need to switch between OMA and MUST when the subband/subchannel is symmetric or not. This entails that, at the base station's side, the scheduler can be ``dummy'', assigning each physical resource block to multiple users without considering whether the subbands/subchannels are symmetric or not. In addition, S-MUST employs the same encoding-decoding mechanism to deal with both symmetric and asymmetric channels. In contrast, as shown in Algorithm 1, OMA+MUST requires to assign a whole piece of physical resource block to a single user when the corresponding subband/subchannel is symmetric, while assigning it to multiple users when the corresponding subband/subchannel is asymmetric. In addition, dynamic MA employs two different encoding-decoding mechanisms, one for symmetric channel, namely, OMA and the other for asymmetric channel, namely, MUST.}

\subsection{Transmission at the BS}

In the proposed S-MUST, the superimposed signal can be generated from the following mapping function
\begin{equation}\label{eqn:generating_x}
x = \lambda \mathcal{W}(v_1, \ldots ,v_L),
\end{equation}
where $v_\ell,\ell\in\{1,...,L\}$ denote the coded symbols of the $\ell$-th UE, e.g., for a $2^m$-ary modulation, ${v_\ell } \triangleq \left[ {{v_{\ell ,1}},....,{v_{\ell ,{m}}}} \right]$ is an $m$-bit binary tuple where $m$ is an integer and $v_{\ell ,t}, t \in\{1,...,m\} $ is the $t$-th bit of $v_\ell$; $\lambda$ is a scaling factor to meet the power constraint; and $\mathcal{W}$ denotes the mapping function. In the following, we will describe three categories of S-MUST, and we will employ $\mathcal{W}_{\text{Cat.}1}$, $\mathcal{W}_{\text{Cat.}2}$ and $\mathcal{W}_{\text{Cat.}3}$ to denote the corresponding mapping functions.

\subsubsection{S-MUST Cat.~1} The mapping function is defined as:
\begin{equation}\label{eqn:cat_1_mapping}
\mathcal{W}_{\text{Cat.}1}\left( v_1,...,v_L \right) \triangleq \sum\limits_{\ell  = 1}^L {{\alpha _\ell }\mathsf{I}\left( {\mathcal{M}_\ell\left( {{v_\ell }} \right)} \right)}  + j\sum\limits_{\ell  = 1}^L {\beta_{\ell} \mathsf{Q}\left( {\mathcal{M}_\ell\left( {{v_\ell }} \right)} \right)},
\end{equation}
where ${\mathcal{M}_\ell\left( \cdot \right)}$ denotes the legacy modulation mapper for each user, e.g., 16-QAM; $\mathsf{I}\left( \cdot \right)$ and $\mathsf{Q}\left( \cdot \right)$ represent the I and Q separation; $\alpha_\ell$ and $\beta_\ell$ denote the I and Q components of the CPAC, respectively, which satisfy the power constraint
\begin{equation}
\mathbb{E}\left[\left| \sum\limits_{\ell  = 1}^L {{\alpha _\ell }\mathsf{I}\left( {\mathcal{M}_\ell\left( {{v_\ell }} \right)} \right)}  + j\sum\limits_{\ell  = 1}^L {\beta_{\ell} \mathsf{Q}\left( {\mathcal{M}_\ell\left( {{v_\ell }} \right)} \right)} \right|^2\right] = {\sum\limits_{\ell  = 1}^L {{\alpha_\ell^2 }}  + \beta_\ell^2  } \le P.	
\end{equation}

\begin{figure}[!htp]
\centering
\includegraphics[width=\figwidth]{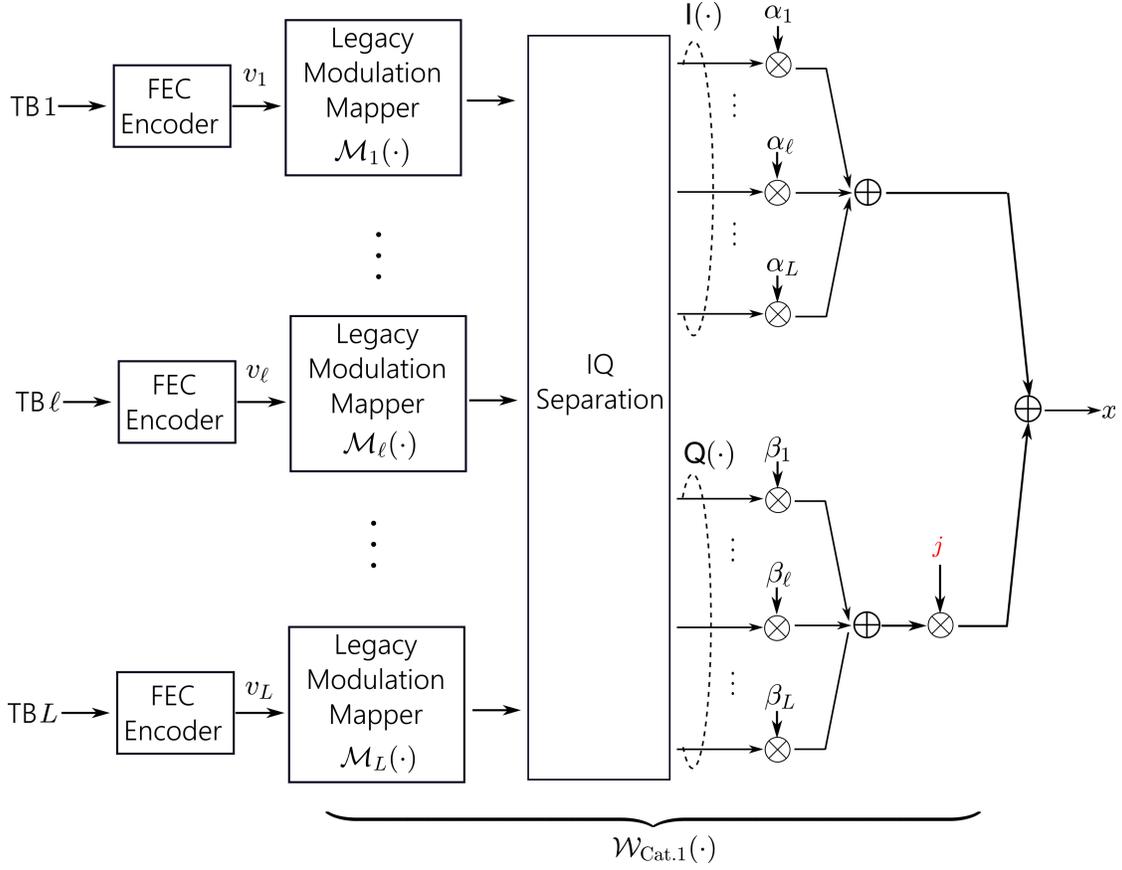}
\caption{Schematic illustration of the proposed S-MUST Cat.~1.}
\label{fig_system_model_S-MUST_catone}
\end{figure}

{A systematic illustration of S-MUST Cat.~1 is shown in Fig.~\ref{fig_system_model_S-MUST_catone}, where transmission block (TB), i.e., data stream, is encoded by forward error correction (FEC) codes and then mapped into legacy constellation. The IQ separation splits I and Q data streams and formulates the mapping function as in \eqref{eqn:cat_1_mapping}.}

\subsubsection{S-MUST Cat.~2} The mapping function is defined as:
\begin{equation}
{\mathcal{W}_{{\text{Cat}}.2}}({v_1}, \ldots ,{v_L}) \triangleq \mathcal{G}\left( {{\mathcal{W}_{{\text{Cat}}.1}}({v_1}, \ldots ,{v_L})} \right),
\end{equation}
where $\mathcal{G}\left(\cdot\right)$ denotes the permutation of Gray labeling.

\subsubsection{S-MUST Cat.~3} {Before we introduce S-MUST Cat.~3, we include some algebra preliminaries as the prelude to S-MUST Cat.~3's mapping function. 

\begin{definition}(Square-free Integers )\label{def_NSI}
An integer is said to be square-free if its prime factorization contains no repeated factors. 
\end{definition}

\begin{definition}
(Modulo operations)  The notation $x\bmod a$ denotes reducing $x \in \mathbb{Z}$ modulo the integer interval $[-a,a)$. That is,
$$x\bmod a=x-b\cdot\left[a-(-a)\right],$$
where $b\in\mathbb{Z}$ is the (unique) integer such that
$$x-b\cdot\left[a-(-a)\right]\in[-a,a).$$
\end{definition}

\begin{lemma} (Chinese Remainder Theorem (CRT) \cite{Hun1974}). Let ${p_1},...,{p_n}$ be relatively prime numbers. For $v_\ell \in \mathbb{Z}_{p_\ell}$, $\ell \in \{1,...,L\}$, there exists a ring isomorphism \cite{Hun1974}:
\[
\mathcal{W}(v_1,...,v_L)=\left(s_1\cdot v_1\cdot \prod_{\ell \neq 1}p_{\ell}+\ldots+s_L\cdot v_L\cdot \prod_{\ell \neq L}p_{\ell}\right)\mod\prod_{\ell}{p_\ell},\]
where $s_1,\ldots,s_L\in\mathbb{Z}$ are such that 
\begin{equation}\label{eqn:ring_iso}
\mathcal{W}(v_1,\ldots,v_L)\mod p_\ell = v_\ell.
\end{equation}
\end{lemma}
We note that $s_1,\ldots,s_L$ can be easily obtained by solving the Bezout's identity and are solely for \eqref{eqn:ring_iso} to hold. For the application to be discussed later, asking \eqref{eqn:ring_iso} may be too much as long as there exists a one-to-one mapping 
so that $v_{\ell}$ can be easily obtained from a simple modulo operation. One such mapping can be obtained by removing $s_1,\ldots,s_L$ to get 
\begin{equation}\label{eqn:CRT2}
\mathcal{W}(v_1,\ldots,v_L)=\left(v_1\cdot \prod_{\ell\neq 1}p_\ell +\ldots+v_L\cdot\prod_{\ell\neq L}p_\ell\right)\mod\prod_\ell p_\ell.
\end{equation}
We note that the first term inside \eqref{eqn:CRT2}, $v_1\cdot\prod_{\ell\neq 1}p_{\ell}$, has every primes except for $p_1$ as its factors (note that $v_1\in\mathbb{Z}_{p_1}$ so cannot be a factor of $p_1$). Moreover, every other term in \eqref{eqn:CRT2} has $p_1$ as its factor. Hence, after $\mod p_1$ operation, only $v_1\cdot (\prod_{\ell\neq 1}p_{\ell})\mod p_1$ remains. Similar reasoning shows leads to 
\begin{equation}\label{eqn:mod_p_l}
\mathcal{W}(v_1,\ldots,v_L)\mod p_\ell = a_\ell\cdot v_\ell \mod p_\ell,
\end{equation}
where $a_\ell = \prod_{\ell'\neq\ell}p_{\ell'}\mod p_\ell$ is independent of $v_{\ell}$. We would like to emphasize that removing $s_1,\ldots,s_L$ allows us to circumvent the complexity of solving Bezout's identity at the transmitter. The price is that each receiver $\ell$ now has to compute $a_{\ell}$, which can be done quite easily.}

\begin{figure}[!htp]
\centering
\includegraphics[width=\figwidth]{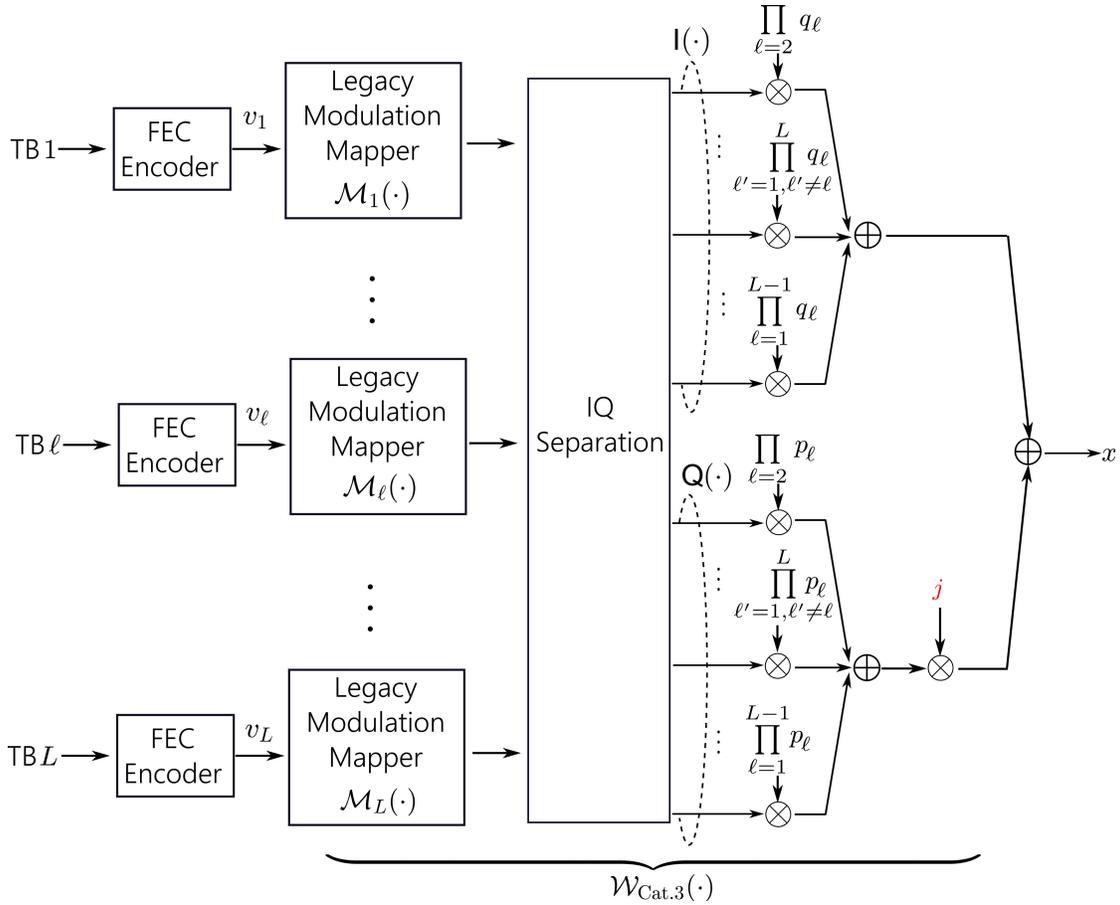}
\caption{Schematic illustration of the proposed S-MUST Cat.~3.}
\label{fig_system_model_S-MUST_catthree}
\end{figure}

In this category, we adopt legacy $2^m$-ary QAM; hence, both the I and Q components become $2^{m/2}$-ary pulse amplitude modulation (PAM). Then  $\alpha_\ell$ and $\beta_\ell$ in S-MUST Cat. 1 are quantized into square-free integers $\hat{\alpha }_\ell$ and $\hat{\beta }_\ell$ which can be factorize into $\hat{\alpha }_\ell = \Pi_{\ell'=1,\ell'\neq \ell}^L {q_{\ell '}}$ and $\hat{\beta }_\ell = \Pi_{\ell'=1,\ell'\neq \ell}^L {p_{\ell '}}$, respectively, where $p_{\ell}>2^{m/2}$ and $q_{\ell}>2^{m/2}$ for $\ell\in\{1,\ldots,L\}$. We then apply the mapping inspired by CRT in \eqref{eqn:CRT2} to get 
\begin{equation}\label{mapping_cat_3}
\begin{aligned}
{\mathcal{W}_{{\text{Cat}}{{.3}}}}\left( {{v_1},...,{v_L}} \right) \triangleq &\left[ {\sum\limits_{\ell  = 1}^L {\left( {\mathsf{I}\left( {\mathcal{M}_\ell\left( {{v_\ell }} \right)} \right) \cdot \mathop \Pi \limits_{\ell ' = 1,\ell ' \ne \ell }^L {q_{\ell '}}} \right)} } \right]\bmod \mathop \Pi \limits_{\ell  = 1}^L {q_\ell }  \\
   &+ j\left[ {\sum\limits_{\ell  = 1}^L {\left( {\mathsf{Q}\left( {\mathcal{M}_\ell\left( {{v_\ell }} \right)} \right) \cdot \mathop \Pi \limits_{\ell ' = 1,\ell ' \ne \ell }^L {p_{\ell '}}} \right)} } \right]\bmod \mathop \Pi \limits_{\ell  = 1}^L {p_\ell },
\end{aligned}
\end{equation}
Based on \eqref{mapping_cat_3}, a systematic illustration of S-MUST Cat.~3 is shown in Fig.~\ref{fig_system_model_S-MUST_catthree}. {The benefit of using this kind of mapping is that M-PIC is feasible at the UE side, which enjoys low system complexity and less overhead. More details are provided in the following subsection B 3).}

\subsection{Reception at the UEs}

Let us consider a BS serving $L$ UEs. The superimposed signal at UE $\ell$ can be written as
\begin{equation}
{y_\ell } = {h_\ell }x + {n_\ell },
\end{equation}
where $x$ is the superposition codeword transmitted by the BS with power $P$; $h_{\ell } $ is the channel coefficient from the BS to UE $\ell $; and ${n_\ell }$ is Gaussian noise with zero mean and variance $2\sigma^2$ per complex dimension; the transmit signal-to-noise ratio at UE $\ell$ is given by
\begin{equation}
\mathsf{SNR}_\ell = \frac{{P\left| {{h_\ell }} \right|}}{2\sigma^2},\forall \ell  \in \left\{ {1,...,L} \right\}.	
\end{equation}

\subsubsection{Detection for S-MUST Cat.~1 and Cat.~2}
By applying channel compensation over the received signal, one can obtain
\begin{equation}
{\tilde y_\ell } = \frac{y_\ell}{h_\ell} = x + {n_{\ell ,{\text{eqv}}}},
\end{equation}
where  ${n_{\ell ,{\text{eqv}}}} \triangleq n_\ell / h_\ell$ is the equivalent noise with variance $2\sigma _{{\rm eqv,}\ell }^{2} =2\sigma ^{2}/\left|h_{\ell } \right|^{2}$.

For simplicity of notation, let ${x_{\ell}} \triangleq {{\mathcal{M}_\ell }\left( {{v_\ell }} \right)}$, ${x_{\ell ,\mathsf{I}}} \triangleq \mathsf{I}\left( x_{\ell}\right)$ and ${x_{\ell ,\mathsf{Q}}} \triangleq \mathsf{Q}\left( x_{\ell} \right)$, $\ell \in\{1,...,L\}$, denote the modulated signal, and its I and Q components, respectively, and let ${{\tilde y}_{\ell ,\mathsf{I}}} \triangleq \mathsf{I}\left( {{{\tilde y}_\ell }} \right)$ and ${{\tilde y}_{\ell ,\mathsf{Q}}} \triangleq \mathsf{Q}\left( {{{\tilde y}_\ell }} \right)$ denote the I and Q components of the received signal, respectively, which can be obtained through IQ separation at the UE. Without loss of generality, let us assume the following channel gain ordering: $\left| {{h_1}} \right| \le\left| {{h_2}} \right| \le\ldots  \le \left| {{h_L}} \right|$. As such, the $\ell$-th UE, $\forall \ell  \in \left\{ {2,...,L} \right\}$ can apply successive interference cancellation (SIC) from the code level of UE $1$ to its own level. Taking the detection of the I component as an example, SIC can be implemented through multistage decoding as follows:
\begin{equation}\label{MSD}
\begin{aligned}
  {{\hat x}_{1,\mathsf{I}}} \approx& \mathop {\arg \max }\limits_{{x_{1,\mathsf{I}}} \in \mathsf{I}\left( {{\mathcal{M}_1}\left( {{\mathbb{Z}_{{2^m}}}} \right)} \right)} {\left| {{{\tilde y}_{1,\mathsf{I}}} - {x_{1,\mathsf{I}}}} \right|^2} \hfill \\
   \vdots&\\
  {{\hat x}_{\ell ,\mathsf{I}}} \approx& \mathop {\arg \max }\limits_{{x_{\ell ,\mathsf{I}}} \in \mathsf{I}\left( {{\mathcal{M}_\ell }\left( {{\mathbb{Z}_{{2^m}}}} \right)} \right)} {\left| {{{\tilde y}_{\ell ,\mathsf{I}}} - {x_{\ell ,\mathsf{I}}} - \sum\limits_{\ell ' = 1}^{\ell  - 1} {{{\hat x}_{\ell ',\mathsf{I}}}} } \right|^2},
\end{aligned}
\end{equation}
where $\hat{x}_{1,\mathsf{I}},...,\hat{x}_{\ell,\mathsf{I}}$ are the recovered I components of the modulated signals. Detection of the Q component can be performed in a similar fashion.

Based on the decoding metric in \eqref{MSD}, the log-likelihood ratios (LLRs) for the $t$-th bit of $v_1,...,v_\ell$ can be represented as follows:
\begin{equation}
\begin{aligned}
  \mathsf{LLR}({v_{1,t}})& \approx \log \frac{{\sum\limits_{{v_{1,t}} = 0} {\exp \left( {\frac{1}{{\sigma _{{\text{eqv}}}^2}}{{\left| {{{\tilde y}_{\ell ,\mathsf{I}}} - {x_{1,\mathsf{I}}}} \right|}^2}} \right)} }}{{\sum\limits_{{v_{1,t}} = 1} {\exp \left( {\frac{1}{{\sigma _{{\text{eqv}}}^2}}{{\left| {{{\tilde y}_{\ell ,\mathsf{I}}} - {x_{1,\mathsf{I}}}} \right|}^2}} \right)} }}, \\
   \vdots&  \\
  \mathsf{LLR}({v_{\ell ,t}}) &\approx \log \frac{{\sum\limits_{{v_{\ell ,t}} = 0} {\exp \left( {\frac{1}{{\sigma _{{\text{eqv}}}^2}}{{\left| {{{\tilde y}_{\ell ,\mathsf{I}}} - {x_{\ell ,\mathsf{I}}} - \sum\limits_{\ell ' = 1}^{\ell  - 1} {{{\hat x}_{\ell ',\mathsf{I}}}} } \right|}^2}} \right)} }}{{\sum\limits_{{v_{\ell ,t}} = 1} {\exp \left( {\frac{1}{{\sigma _{{\text{eqv}}}^2}}{{\left| {{{\tilde y}_{\ell ,\mathsf{I}}} - {x_{\ell ,\mathsf{I}}} - \sum\limits_{\ell ' = 1}^{\ell  - 1} {{{\hat x}_{\ell ',\mathsf{I}}}} } \right|}^2}} \right)} }}, \\
\end{aligned}
\end{equation}
where $\mathsf{LLR}({v_{\ell ,t}})$ is fed to the channel decoder to recover the useful signal.

\subsubsection{Detection for S-MUST Cat. 3} 
\begin{figure}[!htp]
\centering
\includegraphics[width=\figwidth]{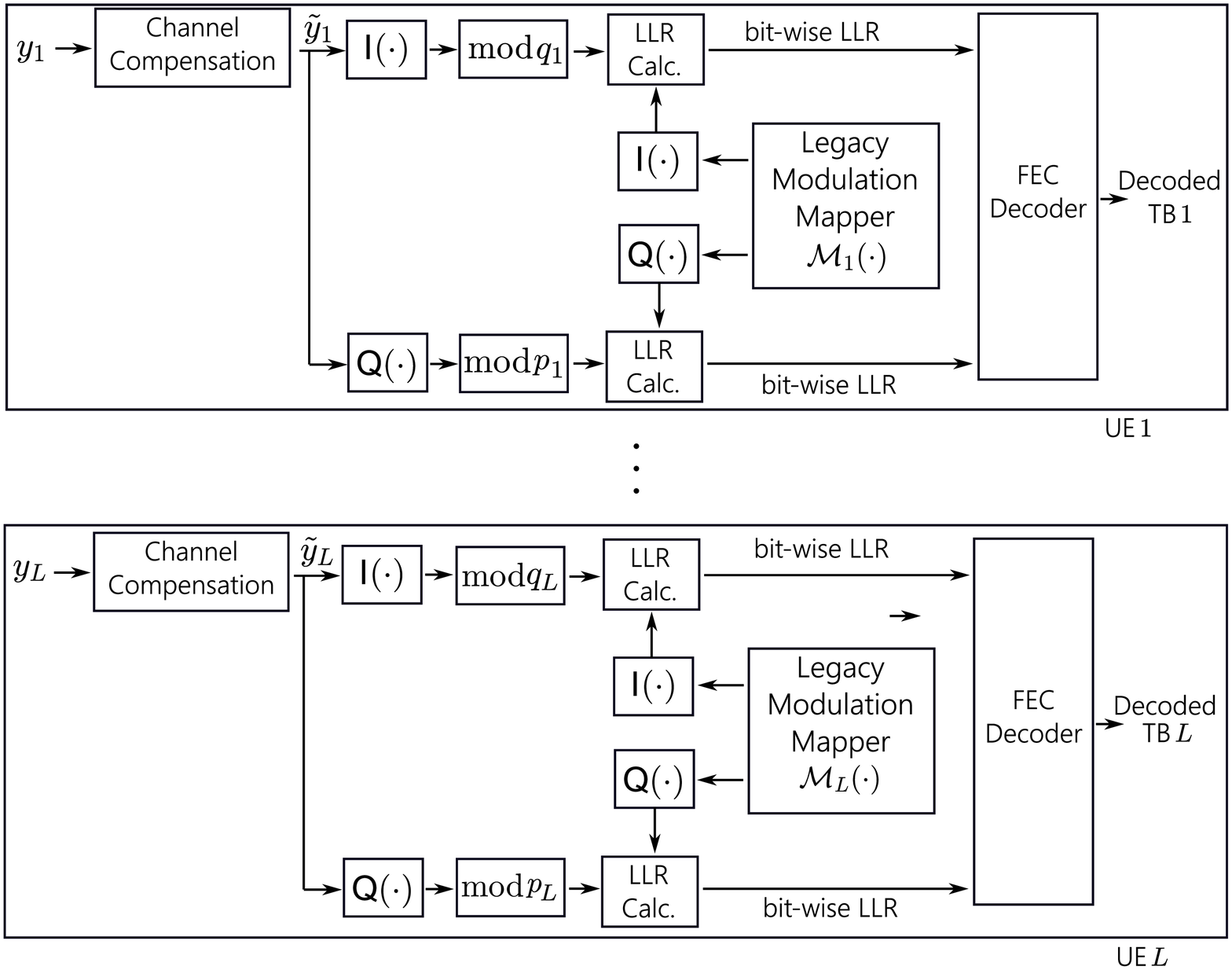}
\caption{M-PIC decoding for S-MUST Cat.~3.}
\label{fig_M_PIC}
\end{figure}

The detection method of S-MUST Cat.~3 is different than that of S-MUST Cat.~1 and Cat.~2 and based on M-PIC. {Let us take UE $\ell$ as an example: as illustrated in Fig.~\ref{fig_M_PIC}, due to the property of CRT described in \eqref{eqn:mod_p_l}, the $\ell $-th code level can be \emph{peeled off} via a modulo operation with respect to ${\theta _\ell }$,  $\forall \ell  \in \left\{ {1,...,L} \right\}$, given by
\begin{equation}
\begin{aligned}
  {{\tilde y}_{\ell ,\mathsf{I},\text{mod}}} &= \left[ {\mathsf{I}\left( {{{\tilde y}_\ell}} \right)} \right]\,\bmod \,{q_\ell },\\
  {{\tilde y}_{\ell ,\mathsf{Q},\text{mod}}} &= \left[ {\mathsf{Q}\left( {{{\tilde y}_\ell}} \right)} \right]\,\bmod \,{p_\ell },
\end{aligned}
\end{equation}
where ${{\tilde y}_{\ell ,\mathsf{I},\text{mod}}}$ and ${{\tilde y}_{\ell ,\mathsf{Q},\text{mod}}}$ denote the I and Q components of the received signal after the modulo operation,} which are fed to the following metric to calculate the bit-wise LLR:
\begin{equation}
\begin{aligned}
\mathsf{LLR}({v_{\ell ,t}}) = \log \frac{{\sum\limits_{{v_{\ell ,t}} = 0} {p\left( {{{\tilde y}_{\ell ,\mathsf{I},\text{mod}}}|{x_{\ell ,\mathsf{I}}}} \right)} }}{{\sum\limits_{{v_{\ell ,t}} = 0} {p\left( {{{\tilde y}_{\ell ,\mathsf{I},\text{mod}}}|{x_{\ell ,\mathsf{I}}}} \right)} }}
= \log \frac{{\sum\limits_{{v_{\ell ,t}} = 0} {\exp \left( {\frac{1}{{\tilde \sigma _{{\text{eqv}}}^2}}{{\left| {{{\tilde y}_{\ell ,\mathsf{I},\text{mod}}} - {x_{\ell ,\mathsf{I}}}} \right|}^2}} \right)} }}{{\sum\limits_{{v_{\ell ,t}} = 0} {\exp \left( {\frac{1}{{\tilde \sigma _{{\text{eqv}}}^2}}{{\left| {{{\tilde y}_{\ell ,\mathsf{I},\text{mod}}} - {x_{\ell ,\mathsf{I}}}} \right|}^2}} \right)} }},
\label{MUST_Cat3_LLR}
\end{aligned}
\end{equation}
where ${\tilde \sigma _{{\text{eqv}}}^2}$ denotes the variance per real dimension of the noise folded through the modulo operation. In high-SNR regime, this can be approximated by the noise variance before the modulo operation. One can observe from (\ref{MUST_Cat3_LLR}) that no SIC decoding is needed, and each user only extracts its desired signals without requiring knowledge of other users' MCS, power level, CQI, and PMI. This proposed M-PIC approach thus significantly reduces the signaling overhead.

\subsection{Power Coefficient Selection}

The selection of the CPACs is crucial for the system performance and to guarantee user fairness. Indeed, the power coefficients should be such that no overlapped points occur on the composite constellation. While the criterion for selecting the power coefficients in \cite{R1-164977} is based on maximizing the minimum Euclidean distance, such approach cannot guarantee optimal user fairness. In contrast, we propose to select the CPACs according to the following maximum-fairness criterion:
\begin{equation}\label{eqn_power_opt}
\begin{aligned}
  \mathop {\max }\limits_{\alpha ,\beta }& \mathop {\min }\limits_{\ell  \in \left\{ {1,...,L} \right\}} I\left( {{Y_\ell };{X_\ell }|{X_{\ell  - 1}},...,{X_1}} \right)\\
  {\text{s}}{\text{.t}}{\text{.}}\quad&{ {\sum\limits_{\ell  = 1}^L {{\alpha _\ell^2 }}  + {{\beta _\ell^2 }} } } \le P, \hfill \\
  &{\alpha _\ell } \ge 0,{\beta _\ell } \ge 0 \hfill
\end{aligned}
\end{equation}
where $I\left( {{Y_\ell };{X_\ell }|{X_{\ell  - 1}},...,{X_1}} \right)$ is the mutual information between received and transmitted signals of the $\ell$-th UE, given that all signals up to $\ell-1$-th have been successfully decoded. One can compute $I\left( {{Y_\ell };{X_\ell }|{X_{\ell  - 1}},...,{X_1}} \right)$ through the chain rule as follows \cite{WacFisHub1999}
\begin{equation}
\begin{aligned}
  I\left( {{Y_\ell };{X_\ell },...,{X_1}} \right) = I\left( {{Y_\ell };{X_1}} \right) + I\left( {{Y_\ell };{X_2}|{X_1}} \right) + I\left( {{Y_\ell };{X_\ell }|{X_{\ell  - 1}},...,{X_1}} \right).
\end{aligned}
\end{equation}
Even though the IQ separation decoding is implemented over the I and Q components separately, the mutual information $I\left( {{Y_\ell };{X_\ell },...,{X_1}} \right)$ takes both the I and Q components into account such that the two degrees of freedom can be jointly exploited.

\begin{figure}[!htp]
\centering
\includegraphics[width=0.5\textwidth]{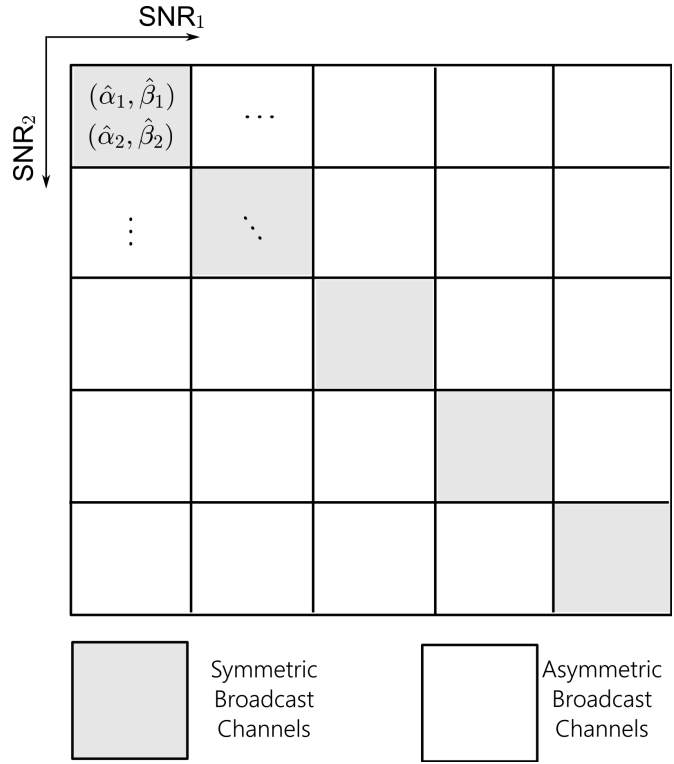}
\caption{Example of LUT for power coefficient selection in the case of two users.}
\label{LUT}
\end{figure}

{As SNR/CQI is the feedback usually adopted in current standardization, one can employ a look-up-table based on the broadcasted SNR/CQIs -- as in equation (17) -- to select the appropriate $q_\ell$. Similar to the method of creating a look-up table (LUT) to select the appropriate modulation and coding scheme in LTE as a function of the SNR, the optimized pair $(\tilde{\alpha}_\ell, \tilde{\beta}_\ell)$ can be stored in an $L$-dimensional LUT at the BS, where each cell corresponds to an unique vector $[\mathsf{SNR}_1,...,\mathsf{SNR}_L]$. Given the feedback $\mathsf{SNR}_\ell, \forall \ell \in \{1,...,l\}$, a BS can select the optimal $(\tilde{\alpha}_\ell, \tilde{\beta}_\ell)$ pair to perform S-MUST transmissions as illustrated in Fig.~\ref{LUT}. As a lightweight solution, in S-MUST Cat.~3 the product of primes can be quantized from the optimized $(\tilde{\alpha}_\ell, \tilde{\beta}_\ell)$ pair. One can adopt the PFA algorithm to find the distinct primes $\tilde{q}_\ell$ and $\tilde{p}_\ell$ on the I and Q components, respectively. Said computations can be performed offline, and one can construct a similar $L$-dimensional LUT to obtain the pair $(p_\ell, q_\ell)$ based on the feedback $[\mathsf{SNR}_1,...,\mathsf{SNR}_L]$.}

\subsection{User Fairness in Symmetric Channels}

The proposed S-MUST is able to multiplex users via superposition transmission without sacrificing user fairness even when they experience similar channel conditions. An example is given as follows, whereas numerical results will be provided in Section~V.

\subsubsection*{Example}
\begin{figure}[!htp]
\centering
\includegraphics[width=\figwidth]{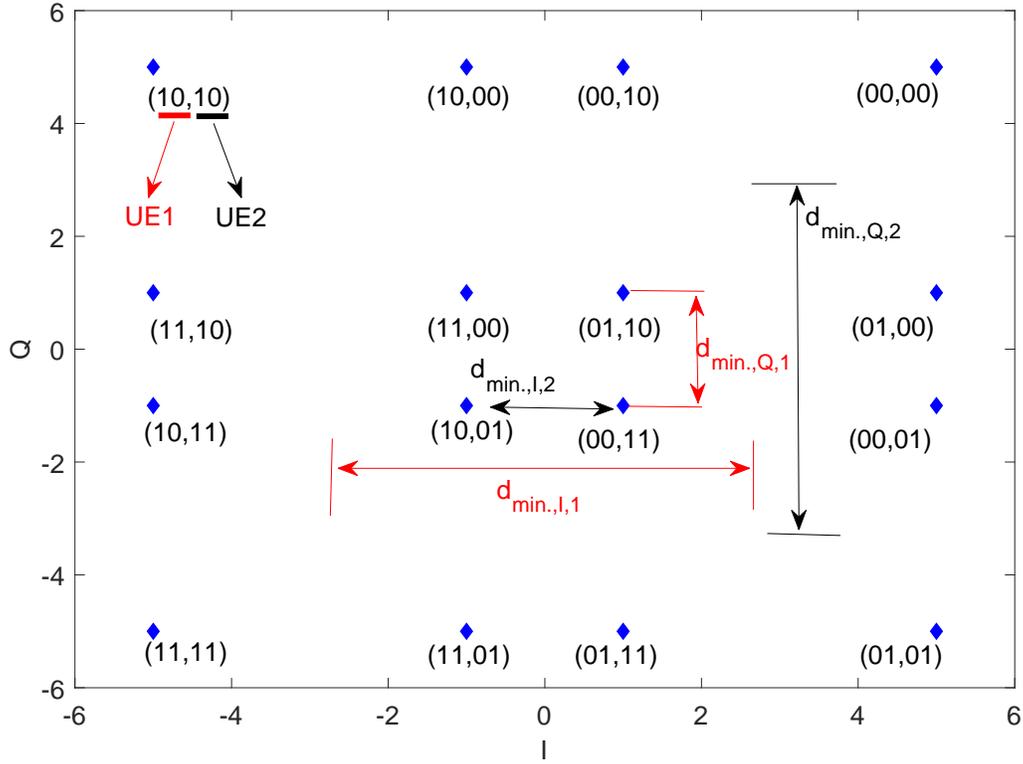}
\caption{Example of composite constellation when both UEs adopt QPSK in a symmetric broadcast channel.}
\label{fig_composite_constellation_QPSK}
\end{figure}

Let us consider two UEs, UE~1 and UE~2, both experiencing similar channel conditions, i.e., $\mathsf{SNR}_1\approx \mathsf{SNR}_2$, and let us assume that both UEs adopt QPSK. {Here is an example: the channel gains are sampled from Rayleigh fading symmetric broadcast channel so that S-MUST gets the CPACs $\alpha_1=2.3$, $\alpha_2=3.11$, $\beta_1=3.01$, and $\beta_2=2.18$ using \eqref{eqn_power_opt} to construct S-MUST Cat.~1 and Cat.~2.}  Then one can quantize said CPACs into $q_1=2$, $q_2=3$, $p_1=3$, and $p_2=2$, obtaining a composite constellation for S-MUST Cat.~3 as the one illustrated in Fig.~\ref{fig_composite_constellation_QPSK}. As S-MUST Cat.~3 adopts IQ separation and M-PIC detection, let $d_{\min,\mathsf{I},\ell}$ and $d_{\min,\mathsf{Q},\ell}$, $\ell \in\{1,2\}$ denote the minimum Euclidean distances of UE $\ell$ on the I and Q components of the composite constellation, respectively. From Fig.~\ref{fig_composite_constellation_QPSK}, we can observe that both UEs have equal error protection in items of Euclidean distances and hence user fairness is guaranteed.

\subsubsection*{Remark}
In the above Example, due to the legacy QPSK constellation mapper, the I component of the composite constellation can be alternatively represented as $\{-5, -1, 1, 5\}$ and $\{$ (1,1), (1,0), (0,1), (0,0) $\}$, where in the latter representation the first and second bits of each pair correspond to UE~1 and UE~2, respectively. Simply applying the modulo operation $\mod3$ -- as per Definition~10 -- at UE~1 over $\{-5, -1, 1, 5\}$ has the desired effect of canceling out the component of UE~2.

\subsection{Proposed Scheduling Algorithm for S-MUST}

Unlike dynamic MA, which opportunistically switches between MUST and OMA, S-MUST is able to provide a unified downlink MA air interface. The latter can significantly reduce the complexity of the PF scheduling operations compared to dynamic MA. Our proposed scheduling algorithm for S-MUST is provided in Algorithm 2.

\begin{algorithm}
\caption*{\textbf{Algorithm 2: Proposed Scheduling Algorithm for S-MUST (Two UEs)}}
\begin{algorithmic}[1]
\State Given the PMI and CQI feedback and the 2-D LUT for optimal CPACs $\left[ {\left( {{{\hat \alpha }_1},{{\hat \alpha }_2}} \right),\left( {{{\hat \beta }_1},{{\hat \beta }_2}} \right)} \right]$;
\State \textit{UE pair selection for S-MUST:}
\For {each sub-band}
 \For {each UE pair $\left( \text{UE}_j,\text{UE}_k \right)$ in $\mathcal{U}$}
 \If {PMI$_j$=PMI$_k$};
 \State $\mathsf{SNR}_j=\frac{P\cdot\text{CQI}_j}{2\sigma^2}$; $\mathsf{SNR}_k=\frac{P\cdot\text{CQI}_k}{2\sigma^2}$;
 \State $\left[ {\left( {{{\hat \alpha }_1},{{\hat \alpha }_2}} \right),\left( {{{\hat \beta }_1},{{\hat \beta }_2}} \right)} \right]=\text{LUT}\left(\mathsf{SNR}_j, \mathsf{SNR}_k\right)$;
 \State calculate the PF$_{j,k}\left(\left[ {\left( {{{\hat \alpha }_1},{{\hat \alpha }_2}} \right),\left( {{{\hat \beta }_1},{{\hat \beta }_2}} \right)} \right]\right)$;
 \Else
 \State continue;
 \EndIf
   \EndFor
\State $\left( {\hat j,\hat k} \right) = \mathop {\arg \max }\limits_{j,k \in \mathcal{U}} \{ {\text{P}}{{\text{F}}_{j,k}}\left( \left[ {\left( {{{\hat \alpha }_1},{{\hat \alpha }_2}} \right),\left( {{{\hat \beta }_1},{{\hat \beta }_2}} \right)} \right] \right)\}$;
\State ${\mathcal{U}_s} \leftarrow {\mathcal{U}_s} \cup \left( {\hat j,\hat k} \right)$.
 \EndFor
\end{algorithmic}
\end{algorithm} 
\section{Design of MIMO-based S-MUST}

\begin{figure*}[!t]
\normalsize
\includegraphics[width=\textwidth]{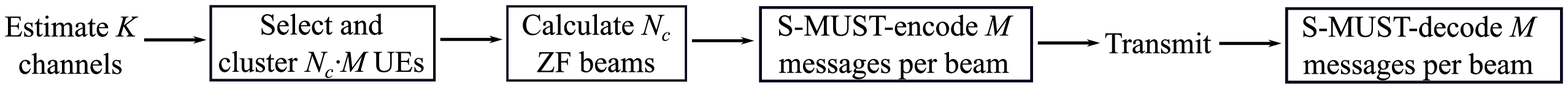}
\caption{Flow chart of the proposed joint design of MIMO and S-MUST.}
\label{fig:Flow_Chart}
\end{figure*}

{In this section, we discuss the joint design of MIMO and S-MUST, including the system model, user clustering, and beamforming design, as shown in subsections A, B, and C, respectively.}

We extend the design of S-MUST to multi-antenna BSs, where the spatial degrees of freedom at each BS can be exploited to create several transmission beams, each carrying signals intended to multiple UEs. Such MIMO-based S-MUST design allows an $N_{\mathrm{t}}$-antenna transmitter to serve $N_{\mathrm{c}} \cdot M$ users on the same PRB. The proposed solution is based on user selection and clustering, zero forcing (ZF) beamforming, and S-MUST encoding/decoding, as illustrated by the flow chart in Fig.~\ref{fig:Flow_Chart}. The remainder of this section will provide a detailed description for each of these building blocks.

\subsection{System Model for MIMO-based S-MUST}

We consider multiuser MISO downlink, where the BS is equipped with $N_{\mathrm{t}}$ antennas, and $K$ is the total number of single-antenna users in a cell. Knowledge of the channels to all $K$ users is assumed to be available at the BS. This can be acquired through orthogonal uplink pilot symbols (for TDD systems) or downlink pilot symbols followed by uplink channel feedback (for FDD systems)\cite{GalCamLopWCNC2018}.

Let $N_{\mathrm{c}} \cdot M=|\mathcal{U}|$ be the number of users scheduled for simultaneous transmission, which are divided into $N_{\mathrm{c}}$ groups or \textit{clusters}, each containing $M \geq 2$ users. We denote by $\mathbf{y}_{n,m}$ the signal received by the $m$-th user in the $n$-th cluster, $m=1,\ldots,M$, $n=1,\ldots,N_{\mathrm{c}}$, given by
\begin{equation}
{y_{n,m}} = \left( {{\mathbf{h}}_{n,m}^{\textrm{H}} {{\mathbf{w}}_n}} \right){x_n} + \underbrace {\sum\limits_{j \ne n} {\left( {{\mathbf{h}}_{n,m}^{\textrm{H}} {{\mathbf{w}}_j}} \right)} {x_j}}_{{e_{n,m}}} + {z_{n,m}},
\label{eqn:rx_signal_I}
\end{equation}
where $(\cdot)^{{\textrm{H}}}$ denotes conjugate transpose; $x_j$ is the superposition codeword transmitted to the $j$-th cluster, generated from \eqref{eqn:generating_x}; $\mathbf{h}_{n,m}^{{\textrm{H}}}$ and $z_{n,m}$ respectively denote the channel vector between the transmitter and the $m$-th user in the $n$-th cluster and the corresponding thermal noise, and $e_{n,m}$ denotes the inter-cluster interference.

The inter-cluster interference $e_{n,m}$ can be reduced by employing linear precoding combined with an efficient user selection and clustering algorithm. Moreover, we note that $x_n$ obtained from \eqref{eqn:generating_x} is the sum of $M$ signals transmitted simultaneously on the same spatial dimension $\mathbf{w}_n$. Therefore, signals intended to users lying in the same cluster create mutual interference. This intra-cluster interference can be removed through an interference cancellation scheme.

\subsection{User Clustering}

\begin{figure}[!t]
\centering
\includegraphics[width=\figwidth]{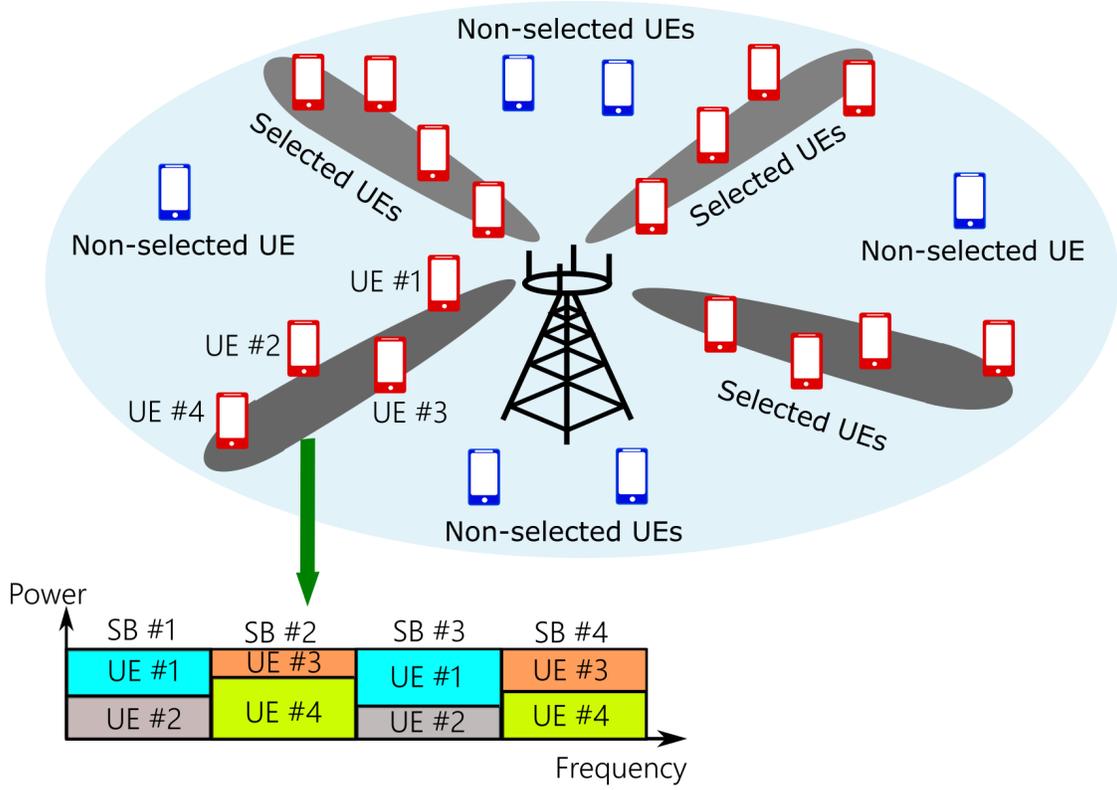}
\caption{Frequency-selective user selection, clustering, and beamforming in MIMO-based S-MUST.}
\label{fig:MIMO_NOMA}
\end{figure}

The proposed clustering algorithm selects $N_{\mathrm{c}} \cdot M$ users out of the $K$ available ones, and groups them into $N_{\mathrm{c}}$ clusters of $M$ users each. Fig.~\ref{fig:MIMO_NOMA} provides an example for the case of $K=20$ available (blue and red) UEs, $N_{\mathrm{t}}=4$ transmit antennas and clusters, and $M=3$ selected (red) UEs per cluster. The algorithm ensures two conditions: (i) that users within the same cluster experience highly correlated channels, and (ii) that users lying in different clusters experience highly uncorrelated channels. The former ensures that all users within a $j$-th cluster receive a strong component of the signal beam intended for that cluster. The latter aims at reducing the inter-cluster interference when paired with ZF precoding. It should be noted that (ii) improves the performance of ZF beamforming with respect to the case when inter-cluster channel correlation is not controlled \cite{YooJSAC06}.

More specifically, the proposed clustering algorithm consists of two phases. In the first phase, one user is selected for each of the $N_{\mathrm{c}}$ clusters, ensuring that the channels $\mathbf{h}_{n,1}$, $n=1,\ldots,N$ of these users, denoted as the \emph{cluster heads}, have significant orthogonal components.\footnote{Some of the operations performed in this phase are similar to the ones in \cite{YooJSAC06} for orthogonal multiuser transmission. However, it should be noted that the algorithm in \cite{YooJSAC06} may fail to find suitable cluster configurations when the numbers $K$ and $N$ are comparable. Another issue with the algorithm in \cite{YooJSAC06} is that it employs an orthogonality threshold whose optimal value is unknown and depends on the system parameters. The two issues above do not occur with the proposed clustering algorithm.} In the second phase of the proposed clustering algorithm, $M-1$ additional users are selected for each cluster, such that all channels $\mathbf{h}_{n,m}$, $m=1,\ldots,M$ of users that lie in the same $n$-th cluster are highly correlated. The two phases of the proposed clustering algorithm are provided in Algorithm~3. Once users have been arranged in clusters, a scheduling algorithm can be employed to obtain the CPACs and generate the superposition transmission. Such procedure is provided in Algorithm~4 and works similarly to what is described in Section~III for the case of single-antenna BSs.

\begin{algorithm}
\caption*{\textbf{Algorithm 3: Proposed Clustering Algorithm for MIMO-based S-MUST} (First Phase)}
\begin{algorithmic}[1]
\State initialize $\mathcal{T}_1 = \{1,\ldots,K\}$
\State initialize $i=1$
\For {each user $k \in \mathcal{T}_1$}
  \State estimate channels $\mathbf{g}_k$
\EndFor
\For {each user $k \in \mathcal{T}_i$} \label{first_loop}
  \State calculate $\tilde{\mathbf{g}}_k$, the component of $\mathbf{g}_k$ orthogonal to the subspace spanned by $\{\mathbf{h}_{1,1},\ldots,\mathbf{h}_{i-1,1}\}$
	\begin{equation*}
\tilde{\mathbf{g}}_k = \mathbf{g}_k - \sum_{j=1}^{i-1} \mathbf{h}_{j,1} \frac{\mathbf{h}_{j,1}^{{\textrm{H}}}\mathbf{g}_k}{\|\mathbf{h}_{j,1}\|^2}
\end{equation*}
(when $i=1$, this implies $\tilde{\mathbf{g}}_k = \mathbf{g}_k$)
\EndFor
\State select the first user for the $i$-th cluster as
\begin{equation*}
\pi(i) = \argmax_{k \in \mathcal{T}_i} \|\tilde{\mathbf{g}}_k\|
\end{equation*}
\State $\mathcal{U}_i = \left\{\pi(i)\right\}$
\State $\mathcal{T}_{i+1} = \mathcal{T}_{i} \backslash \left\{\pi(i)\right\}$
\State $\mathbf{h}_{i,1} = \mathbf{g}_{\pi(i)}$
\State $i \leftarrow i+1$
\If {$\mathcal{T}_{i+1}$ is nonempty and $i\leq N_{\mathrm{c}}$}
\State go to line \ref{first_loop}
\Else
  \State the first phase is completed, go to line \ref{second_phase} 
\EndIf \label{end_first_phase}
\end{algorithmic}
\end{algorithm}

\begin{algorithm}
\caption*{\textbf{Algorithm 3: Proposed Clustering Algorithm for MIMO-based S-MUST} (Second Phase)}
\begin{algorithmic}[1]
\makeatletter
\setcounter{ALG@line}{18}
\State reconsider all remaining users \label{second_phase}
\begin{equation*}
\mathcal{T} = \{1,\ldots,K\}\backslash \cup_{j=1}^{N_{\mathrm{c}}} \mathcal{U}_j	
\end{equation*}
\State initialize $m = 2$	
\For {each user $k \in \mathcal{T}$}
	\For {$n=1,\ldots,N_{\mathrm{c}}$}
	\State calculate $\bar{{g}}_{k,n}$, the correlation between $\mathbf{g}_k$ and $\mathbf{h}_{n,1}$
	\begin{equation*}
\bar{{g}}_{k,n} = \frac{|\mathbf{h}_{n,1}^{{\textrm{H}}}\mathbf{g}_k|}{\|\mathbf{h}_{n,1}\|\|\mathbf{g}_{k}\|}
\end{equation*}
\EndFor
\EndFor
\For {$n=1,\ldots,N_{\mathrm{c}}$} \label{second_loop}
	\State select the most correlated user as
\begin{equation*}
\pi_n(m) = \argmax_{k \in \mathcal{T}} |\bar{{g}}_{k,n}|
\end{equation*}
\State $\mathcal{U}_n \leftarrow \mathcal{U}_n \cup \left\{\pi_n(m)\right\}$
\State $\mathbf{h}_{n,m} = \mathbf{g}_{\pi_n(m)}$
\State $\mathcal{T} = \mathcal{T}\backslash\pi_n(m)$
\EndFor
\State $m \leftarrow m+1$
\If {$m\leq M$}
	\State go to line \ref{second_loop}
\Else
\State the second phase is completed, go to line \ref{third_phase}
\EndIf
\end{algorithmic}
\end{algorithm}

\begin{algorithm}
\caption*{\textbf{Algorithm~4: Proposed Scheduling Algorithm for MIMO-based S-MUST}}
\begin{algorithmic}[1]
\For {$n=1,\ldots,N_{\mathrm{c}}$} \label{third_phase}
\For {each sub-band}

 \For {each UE pair $\left( \text{UE}_j,\text{UE}_k \right)$ in $\mathcal{U}_n$}
 \State $\mathsf{SNR}_j=\frac{P\cdot\text{CQI}_j}{2\sigma^2}$; $\mathsf{SNR}_k=\frac{P\cdot\text{CQI}_k}{2\sigma^2}$;
 \State $\left[ {\left( {{{\hat \alpha }_1},{{\hat \alpha }_2}} \right),\left( {{{\hat \beta }_1},{{\hat \beta }_2}} \right)} \right]=\text{LUT}\left(\mathsf{SNR}_j, \mathsf{SNR}_k\right)$;
 \State calculate the PF$_{j,k}\left(\left[ {\left( {{{\hat \alpha }_1},{{\hat \alpha }_2}} \right),\left( {{{\hat \beta }_1},{{\hat \beta }_2}} \right)} \right]\right)$;

   \EndFor
\State $\left( {\hat j,\hat k} \right) = \mathop {\arg \max }\limits_{j,k \in \mathcal{U}_n} \{ {\text{P}}{{\text{F}}_{j,k}}\left( \left[ {\left( {{{\hat \alpha }_1},{{\hat \alpha }_2}} \right),\left( {{{\hat \beta }_1},{{\hat \beta }_2}} \right)} \right] \right)\}$;
\State ${\mathcal{U}_{s,n}} \leftarrow {\mathcal{U}_{s,n}} \cup \left( {\hat j,\hat k} \right)$.
 \EndFor
 \EndFor
\end{algorithmic}
\end{algorithm}

\subsection{Zero Forcing Beamforming}

After UEs have been selected and clustered, each BS adopts ZF beamforming for the simultaneous transmission of signals to different clusters. Zero forcing beamforming is of particular interest because it is a linear scheme with low-complexity implementation, and because it can control the amount of interference across clusters \cite{Peel05,GeraciJSAC,YanGerQueTSP2016}. In our proposed MIMO-based S-MUST design, each BS stacks up the $N_{\mathrm{c}}$ channels to the selected cluster heads in the following matrix
\begin{equation}
\mathbf{H}=[\mathbf{h}_{1,1}^{\textrm{T}},\ldots,\mathbf{h}_{N_{\mathrm{c}},1}^{\textrm{T}}]
\end{equation}
and calculates the beamforming vectors $\mathbf{w}_n$, $n=1,\ldots,N_{\mathrm{c}}$, as follows
\begin{equation}
\mathbf{w}_n = \frac{1}{\sqrt{\gamma}}\mathbf{h}_{n,1}^{{\textrm{H}}} \left(\mathbf{H} \mathbf{H}^{{\textrm{H}}} \right)^{-1},
\end{equation}
where $(\cdot)^{\textrm{T}}$ denotes transpose and $\gamma = \textrm{tr}\{\mathbf{H}^{{\textrm{H}}}\mathbf{H}(\mathbf{H}\mathbf{H}^{{\textrm{H}}})^{-2}\}$ is a power normalization constant. We note that under ZF beamforming the following condition holds
\begin{equation}
\mathbf{h}_{n,1} \mathbf{w}_j = 0 \quad \forall j \neq n,
\end{equation}
therefore cluster heads do not receive any inter-cluster interference. However, all remaining users in each cluster do receive inter-cluster interference, since
\begin{equation}
\mathbf{h}_{n,m} \mathbf{w}_j \neq 0 \quad \textrm{if} \enspace m \neq 1,
\end{equation}
and such interference is treated as noise and dealt with by the S-MUST decoder.

\subsection{Encoding and Decoding}

On each beam formed by the ZF precoder, superposition transmission and reception is performed according to the S-MUST encoding and decoding procedures described in Section III.
\section{Performance Evaluation}

In this section, we evaluate the performance of the proposed S-MUST scheme. A detailed list of the simulation parameters is provided in Table~\ref{parameters}.

\begin{table}[!htp]
\centering
\small
\caption{Simulation parameters.}
\begin{tabular}{|c|c|}
\hline
{Cellular Layout}  & {Hexagonal, wrapped around} \\ \hline
{Topology} & {$7$ sites (no sectorization)} \\ \hline
{Bandwidth} & {$10$~Mhz} \\ \hline
{Tx Antenna No.} & {$1$ or $2$ (omni-directional)} \\ \hline
{Rx Antenna No.} & {$1$ (omni-directional)} \\ \hline
{No. of UEs per cell} & {$150$ (full-buffer traffic model)} \\ \hline
{BS inter-site distance} & {$500$~m} \\ \hline
{BS Tx power} & {$46$~dBm} \\ \hline
{Thermal noise density} & {$-174$~dBm/Hz} \\ \hline
{Rx noise figure} & {$5$~dB} \\ \hline
{Path loss model} & {$128.1 + 37.6\log_{10}(D)$, $D$ in km} \\ \hline
{Fast fading} & {i.i.d. Rayleigh fading} \\ \hline
\end{tabular}
\label{parameters}
\end{table}

\subsection{Performance of S-MUST}

In what follows, QPSK or 16-QAM are adopted as the component constellation for each user, which form 16-QAM or 256-QAM composite constellations, respectively.

\begin{figure}[!htp]
\centering
\includegraphics[width=\figwidth]{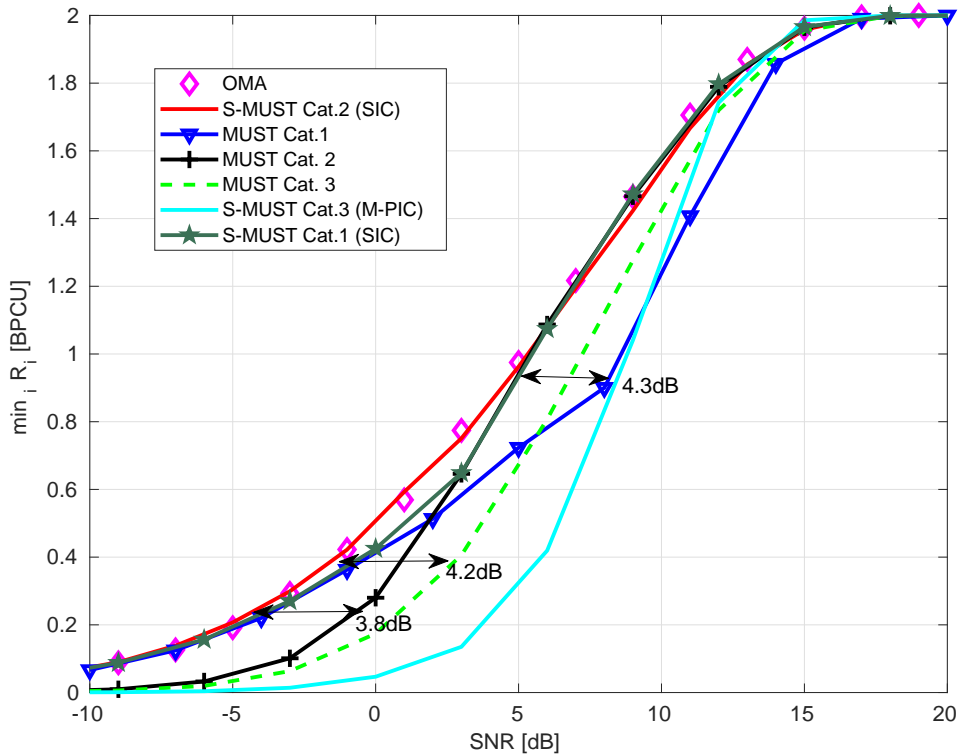}
\caption{User fairness comparison in symmetrical broadcast channels under QPSK.}
\label{fig_fairness_QPSK}
\end{figure}

Fig.~\ref{fig_fairness_QPSK} compares the user fairness of several schemes in symmetric broadcast channels in terms of minimum bits per channel use (BPCU) -- i.e., those of the worst user -- versus SNR, where $h_1=h_2=1$ and each user adopts QPSK modulation. {We used SIC for S-MUST Cat. 1 and 2 and MUST Cat. 1-3 to generate the results in Fig 9, and M-PIC for S-MUST Cat. 3.}  We can observe that: (i) S-MUST outperforms MUST Cat.~1 in regimes of moderate SNR, exhibiting a 4.3~dB enhancement; (ii) S-MUST Cat. 2 outperforms MUST Cat.~2 in low-SNR regime with a 3.8~dB enhancement; (iii) S-MUST outperforms MUST Cat.~3 in low-SNR regime with a 4.2~dB enhancement; (iv) {S-MUST Cat. 1 and 2 achieve nearly equal performance while S-MUST cat.1 is slightly worse;} (v) {S-MUST Cat. 3 performance is worse than OMA and all SIC-based schemes, as M-PIC is a sub-optimal decoder, while it enjoys lower complexity and less overhead;} and (vi) S-MUST almost achieves the same user fairness as OMA, i.e., equal user rates.

\begin{figure}[!htp]
\centering
\includegraphics[width=\figwidth]{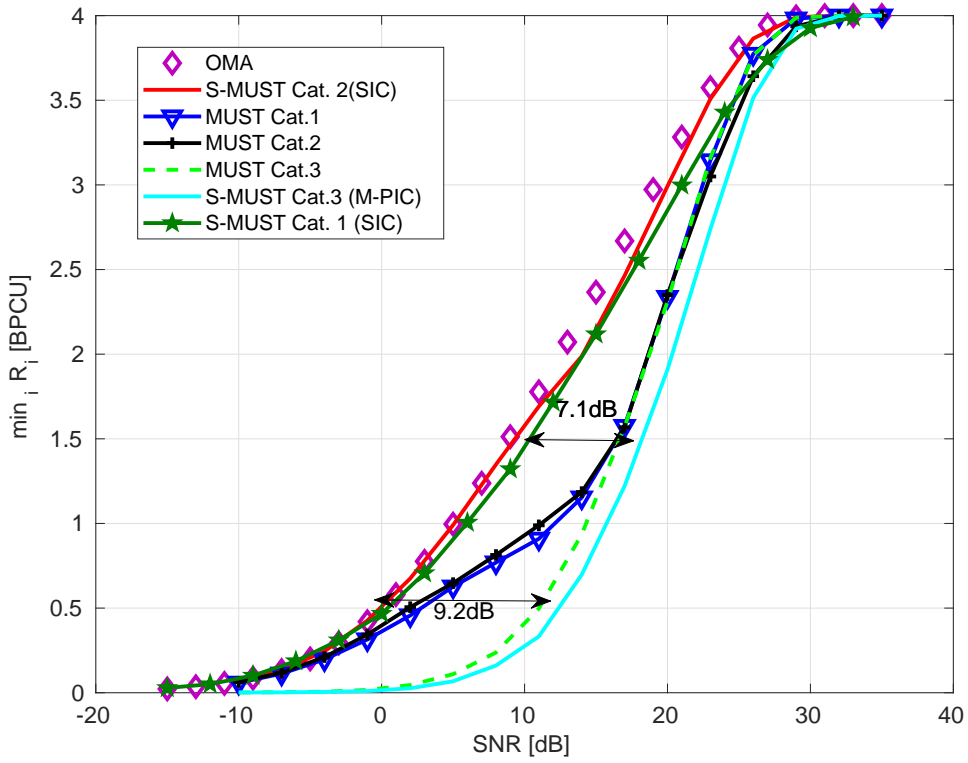}
\caption{User fairness comparison in symmetrical broadcast channels under 16-QAM.}
\label{fig_fairness_16QAM}
\end{figure}

Fig.~\ref{fig_fairness_16QAM} provides a similar performance comparison for the case where each user adopts a 16-QAM modulation. Similar observations can be made: (i) S-MUST Cat. 2 outperforms MUST Cat.~1 in moderate-SNR regime with a 7.1~dB enhancement; (ii) S-MUST Cat. 2 outperforms MUST Cat.~2 in low-SNR regime with a 6.3~dB enhancement; (iii) S-MUST Cat. 2 outperforms MUST Cat.~3 in low-SNR regime with a 9.2~dB enhancement; (iv) {S-MUST Cat. 1 and 2 achieve nearly equal performance;} (v)  {S-MUST cat. 3 performance is worse than OMA and all SIC based schemes as M-PIC is sub-optimal decoder while it enjoys the lowest complexity and less overhead;}  (vi) S-MUST Cat. 1 and 2 almost achieve the same user fairness as OMA.

\subsection{Performance of MIMO-based S-MUST}

\begin{figure}[!htp]
\centering
\includegraphics[width=0.9\figwidth]{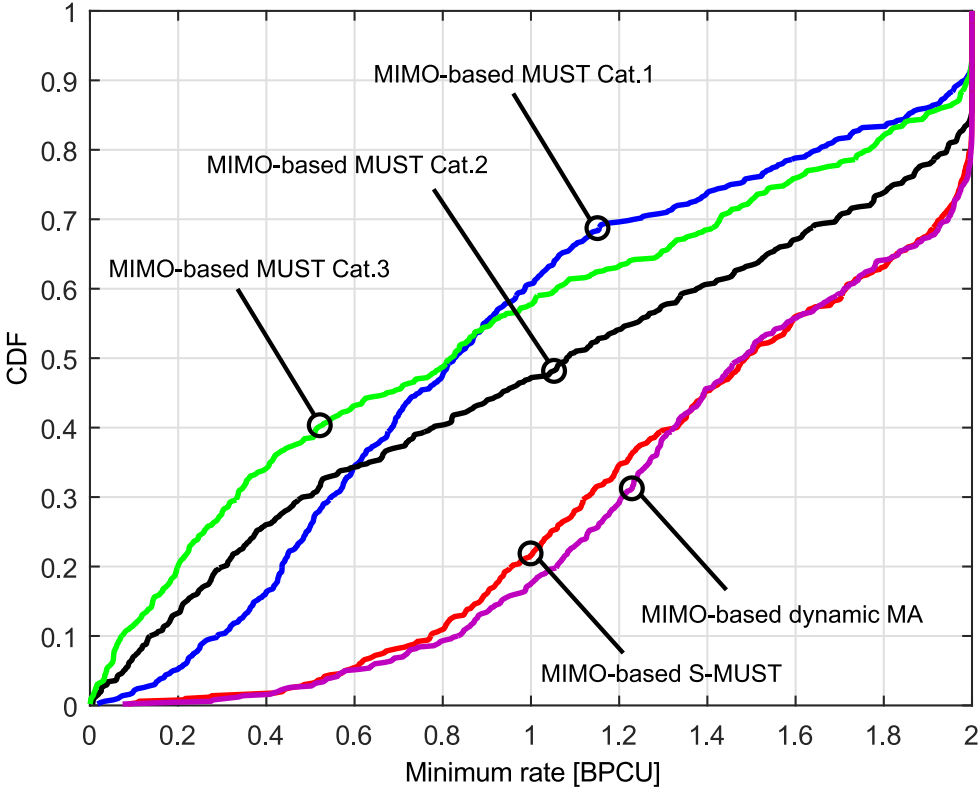}
\caption{User fairness comparison in symmetrical multi-antenna broadcast channels under QPSK.}
\label{fig_fairness_QPSK_MIMO}
\end{figure}

\begin{figure}[!htp]
\centering
\includegraphics[width=0.93\figwidth]{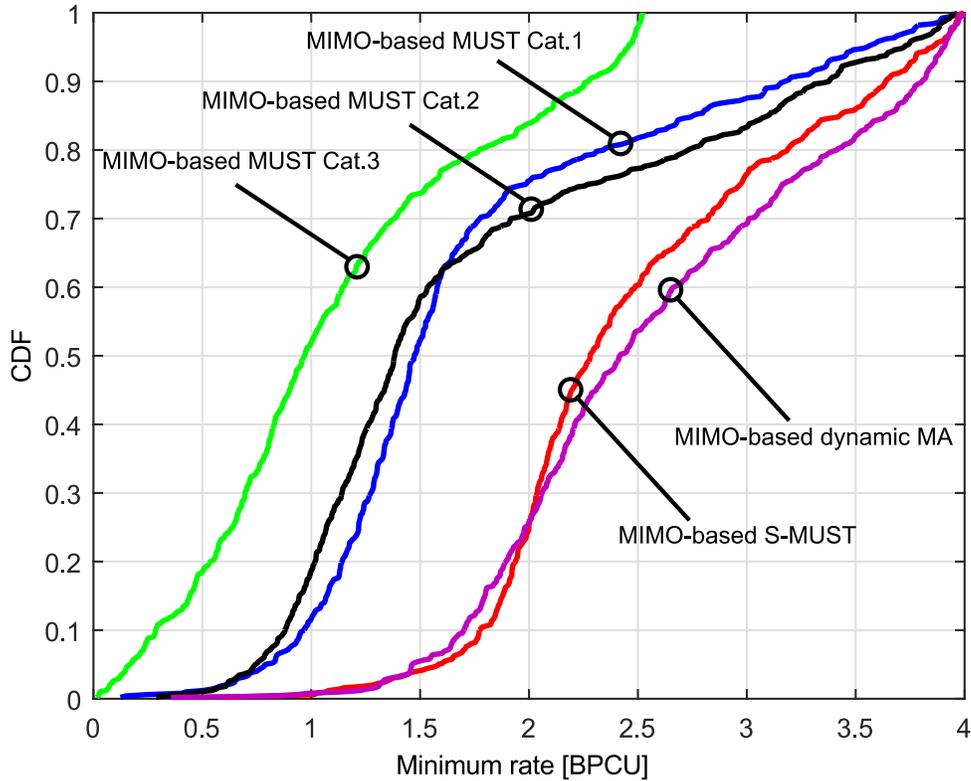}
\caption{User fairness comparison in symmetrical multi-antenna broadcast channels under 16-QAM.}
\label{fig_fairness_16QAM_MIMO}
\end{figure}

In the following, we evaluate the performance of the proposed MIMO-based S-MUST design, by comparing it to MIMO-based designs of conventional MUST, and to a MIMO-based dynamic MA approach where OMA and MUST are opportunistically alternated. In what follows, each BS is equipped with 2 antennas, each UE is equipped with a single antenna, and 4 UEs share the same PRB. Fig.~\ref{fig_fairness_QPSK_MIMO} and Fig.~\ref{fig_fairness_16QAM_MIMO} show the cumulative distribution function (CDF) of the minimum rate -- i.e. that of the worst user --, where QPSK or 16-QAM are adopted as the component constellation for each user, respectively, yielding 16-QAM or 256-QAM composite constellations. In both cases, one can observe that MIMO-based S-MUST achieves almost equal fairness performance as the one of dynamic MA. Moreover, MIMO-based S-MUST outperforms MIMO-MUST Cat.~1, Cat.~2, and Cat.~3 across the whole rate region. In particular, for the $5\%$-worst rate (bottom-left region of the curves, representing the cell edge), MIMO-based S-MUST can provide a two-to-three-fold rate gain.

\section{Conclusion}
We proposed a new downlink multiuser superposition transmission scheme for future 5G cellular networks, which we denoted \emph{structured multiuser superposition transmission} (S-MUST). In S-MUST, we apply complex power allocation coefficients (CPACs) over users' legacy constellations to generate a composite constellation. Said CPACs offer an extra degree of freedom for multiplexing users while ensuring that fairness is guaranteed even for symmetric broadcast channels. The newly proposed paradigm of superposition coding provides a unified multiple access air interface, and allows simple parallel decoding based on IQ separation, CPAC quantization, and modulo operations. We also devised suitable scheduling operations for S-MUST, and designed a MIMO-based version of S-MUST for multi-antenna BSs. We demonstrated that the proposed S-MUST design achieves better user fairness compared with conventional MUST, while exhibiting lower complexity compared to dynamic MA.
\ifCLASSOPTIONcaptionsoff
  \newpage
\fi
\bibliographystyle{IEEEtran}
\bibliography{Bib_Gio_Dong}
\end{document}